\documentclass[sigconf,11pt]{acmart}

\renewcommand\footnotetextcopyrightpermission[1]{}
\usepackage{listings}
\usepackage{xcolor}
\usepackage{float}

\AtBeginDocument{%
  \providecommand\BibTeX{{%
    \normalfont B\kern-0.5em{\scshape i\kern-0.25em b}\kern-0.8em\TeX}}}

\setcopyright{none}
\copyrightyear{}
\acmYear{}
\acmDOI{}

\acmPrice{}
\acmISBN{}

\begin{document}

\title{\Large Prediction of Spotify Chart Success Using Audio and Streaming Features}

\author{\normalsize Cabansag, Ian Jacob and Ntegeka, Paul}


\maketitle 
\pagestyle{plain} 
\section{Abstract}

Spotify’s streaming charts offer a real-time lens into music popularity, driving discovery, playlists, and even revenue potential. Understanding what influences a song’s rise on these charts—especially early on—can guide marketing efforts, investment decisions, and even artistic direction. In this project, we developed a classification pipeline to predict a song’s chart success based on its musical characteristics and early engagement data. Using all 2024 U.S. Top 200 Spotify Daily Charts and the Spotify Web API, we built a dataset containing both metadata and audio features for 14,639 unique songs.

The project was structured in two phases. First, we benchmarked four models: Logistic Regression, K Nearest Neighbors, Random Forest, and XGBoost—using a standard train-test split. In the second phase, we incorporated cross-validation, hyperparameter tuning, and detailed class-level evaluation to ensure robustness. Tree-based models consistently outperformed the rest, with Random Forest and XGBoost achieving macro F1-scores near 0.95 and accuracy around 97\%.

Even when stream count and rank history were excluded, models trained solely on audio attributes retained predictive power. These findings validate the potential of audio-based modeling in A\&R scouting, playlist optimization, and hit forecasting—long before a track reaches critical mass.

\section{Introduction}

Streaming platforms have transformed how music is distributed and consumed, and Spotify sits at the forefront of this ecosystem. Their Top 200 Daily Charts serve not only as a barometer of public taste, but also as a powerful determinant of a song’s trajectory—impacting visibility, playlist placement, and monetization opportunities. For artists and labels alike, anticipating whether a song will chart—and how high—has become both an art and a science.

This project set out to answer whether machine learning models, trained on a combination of musical and engagement features, could classify a song’s peak chart rank. Specifically, we aimed to predict whether a song would land in the Top 10, Ranks 11–50, or Ranks 51–100. We pursued this through a rigorous modeling framework that began with simple benchmark models and evolved into a robust, cross-validated system based on feedback. Our goal was not only predictive accuracy but also model transparency, balance across class labels, and interpretability.

\section{Purpose}

At its core, this project was motivated by a desire to understand how early signals in a song’s lifecycle—especially those available immediately upon release—might predict long-term chart success. Could a song’s composition alone, even without knowing how many people had streamed it, hint at whether it would become a hit?

Three key objectives guided my approach:

\begin{enumerate}
    \item \textbf{Feasibility: }Determine whether audio-only models could predict peak chart performance in the absence of stream counts or rank metadata.
    \item \textbf{Feature Impact: }Evaluate the influence of metadata like stream count and previous rank when included or excluded from the model.
    \item \textbf{Model Rigor:} Compare classifiers not just by accuracy but by macro-averaged F1, cross-validation variance, and class-specific performance.
\end{enumerate}

Initially, we approached the problem pragmatically, focusing on a simple holdout split and base model performance. But feedback from early reviewers helped pivot the project into a more rigorous direction. We learned to consider model variance, tuning sensitivity, and fairness across classes—insights that proved essential in developing models that are both powerful and trustworthy.

\section{Data and Feature Engineering}

\subsection{Data Collection and Preparation}

To build the dataset, we combined two primary data sources: (1) the publicly available Spotify U.S. Top 200 Daily Charts, and (2) audio feature data retrieved from Spotify’s Web API.

The first stage involved downloading all daily chart files for the year 2024 from SpotifyCharts.com. These files included song title, artist, Spotify URI, daily rank, stream count, and date. Using our R script music charts dataset usa, we concatenated over 360 CSV files into a single panel structure and extracted the peak rank that each track achieved during the year. This became the target label in the classification task.

The second stage required enriching the dataset with \textbf{track-level musical feature}s. To do this, we used the Spotify Web API via the R package tinyspotifyr, which wraps Spotify’s GET endpoints with helpful token management and pagination functionality.

Spotify’s API \cite{SpotifyAPI2024} allows querying up to 100 track IDs per call, so we chunked the list of unique track URIs into batches of 100. We implemented a batching function that dynamically created request URLs, handled query-string encoding, and used a loop to paginate through all entries.

To respect Spotify’s API rate limits, we followed a conservative strategy of issuing at most 2 requests per second (well under the stated 10 requests/second soft limit). We also implemented a retry mechanism using exponential backoff, in case a 429 Too Many Requests response was encountered. This ensured that the process could run overnight without manual supervision and without being blocked.

In total, over 14,600 unique tracks were enriched with the following 13 audio features:

\begin{itemize}
    \item danceability, energy, valence, tempo, acousticness, loudness, instrumentalness, speechiness, liveness, key, mode, duration\_ms, and time\_signature.
\end{itemize}

An R script performed the data joining, merging the chart and feature datasets on Spotify’s unique track URI. We also ensured that each track had exactly one entry, retaining only the first appearance in cases of duplicate peak ranks. Missing values were rare (<0.2\%) and were imputed using the mean of the training set during preprocessing.

By the end of this process, we had created a high-integrity dataset that captured both quantitative musical structure and real-world streaming performance. This combination made it ideal for exploring predictive modeling applications in the music industry.

\subsection{Feature Structure and Target Labeling}

We grouped each song into one of three outcome classes based on its peak chart position for the 2024 calendar year. This decision was grounded in how the music industry interprets ranking tiers. Songs that reach the Top 10 often benefit from large-scale promotion, frequent radio rotation, and premium playlist placements on Spotify’s editorial ecosystem. Tracks peaking in Ranks 11 to 50 represent rising or mid-level performers—often propelled by organic virality, targeted marketing, or loyal fanbases. Songs that fall within Ranks 51 to 100 are typically either early in their promotional cycle or plateaued in exposure.

To implement this, We labeled each track using:

\begin{itemize}
    \item \textbf{Top 10: }Songs that peaked between Ranks 1 and 10 
    \item \textbf{11–50: }Songs that peaked between Ranks 11 and 50 
    \item \textbf{51–100: }Songs that peaked between Ranks 51 and 100 
\end{itemize}

An analysis of class distribution revealed a notable imbalance: only 732 songs reached the Top 10 class, while the 11–50 and 51–100 categories contained 2,928 and 10,979 tracks, respectively. This imbalance had important implications for both training and evaluation. In particular, it reinforced the need to use macro-averaged F1-score, which treats each class equally regardless of frequency. It also prompted the use of stratified splitting in both the 80/20 holdout split and cross-validation, ensuring proportional representation of each class.

Each song’s features were organized into two categories: metadata (e.g., streams, previous\_rank, peak\_rank, days\_on\_chart) and audio features extracted from the Spotify Web API. All numerical columns were scaled using StandardScaler to support models sensitive to feature magnitude, such as Logistic Regression and K-Nearest Neighbors. Categorical columns, such as mode and key, were encoded numerically.

To test the value of musical composition alone, we also created a filtered dataset by excluding metadata features. This mimicked real-world scenarios where record labels may want to evaluate unreleased songs prior to public availability. The filtered version retained only the 13 audio features and provided a more stringent test of whether a song’s structure—not its early exposure—could signal commercial potential.

These dual versions of the dataset—full and filtered, allowed for a fair comparison across modeling scenarios and helped isolate the specific contributions of musical features.

\section{Exploratory Data Analysis (EDA)}
\subsection{Audio Feature Distributions and Outliers}

Before initiating model training, we conducted exploratory analysis of feature distributions and outlier behavior to understand the underlying data landscape. The streams variable, as expected, exhibited a strong right-skew, with Top 10 tracks frequently logging millions of daily plays, especially in the early days of release. We considered log-transforming this feature to reduce skew, but ultimately kept the raw version, as tree-based models are unaffected by monotonic transformations and would preserve interpretability better without transformation artifacts.

Features like valence, tempo, and loudness followed roughly normal distributions, albeit with slight skews. Tempo had interesting clustering behavior: high-tempo songs often fell in Top 10 and 11–50 classes, supporting the idea that more energetic tracks perform better commercially. Meanwhile, loudness and energy were highly correlated ($r \approx 0.71$), likely reflecting modern production norms, where highly compressed and louder mixes tend to accompany up-tempo music.

In contrast, speechiness and instrumentalness were bimodal, with clusters near 0 and 1. This pattern pointed to genre-driven differences. For instance, tracks with high speechiness were predominantly hip-hop, often trending in Top 50. Tracks with high instrumentalness, such as classical or ambient music, rarely broke into Top 10, underscoring genre preferences in mainstream streaming behavior.

We carefully evaluated how to handle extreme values, particularly in the streams feature. While techniques like winsorization or z-score clipping could reduce their influence, we chose to retain all outliers. These data points are often the most interesting—representing explosive hits like Olivia Rodrigo’s “vampire” or Drake’s surprise releases. Removing them would suppress key examples that models need to learn how to differentiate. Moreover, Random Forest and XGBoost are inherently robust to such skew due to their hierarchical split logic: they simply partition outliers into terminal leaf nodes without letting them skew broader decisions.

No feature was discarded during preprocessing. Even weakly correlated variables were retained on the hypothesis that they may hold interactions or non-linear relationships that linear correlation cannot reveal. This decision proved correct—several lower-correlation features became highly ranked in importance after training.

Together, these findings confirmed that audio features encode important genre and compositional signals, which—when paired with metadata or on their own—can drive meaningful predictive performance.

\subsection{Correlation and Feature Redundancy}

To understand relationships between variables, we computed the pairwise correlation matrix. The highest observed correlation was between loudness and energy (r = 0.71), followed by moderate relationships between acousticness and instrumentalness. However, no two features exhibited multicollinearity levels high enough to justify removal.

\begin{center}
\includegraphics[width=1\linewidth]{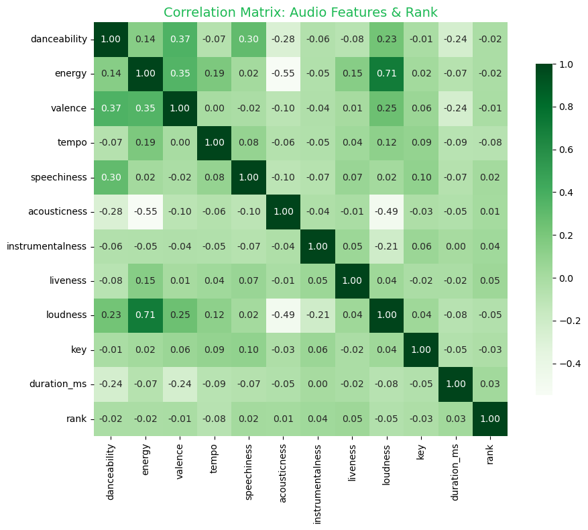}
\end{center}

This matrix highlighted that most features had weak correlation with chart rank individually, implying that combinations and interactions—not linear relationships, would be needed for accurate classification. This insight influenced my choice to favor tree-based models over purely linear ones.

\subsection{Rank-Class Relationships and Early Stream Counts}

The scatter plot below visualizes how stream count varied by rank and rank class. As anticipated, Top 10 tracks had the highest stream counts, though there was overlap across classes. Some songs outside the Top 10 had strong early performance but plateaued; others rose more gradually.

\begin{center}
\includegraphics[width=1\linewidth]{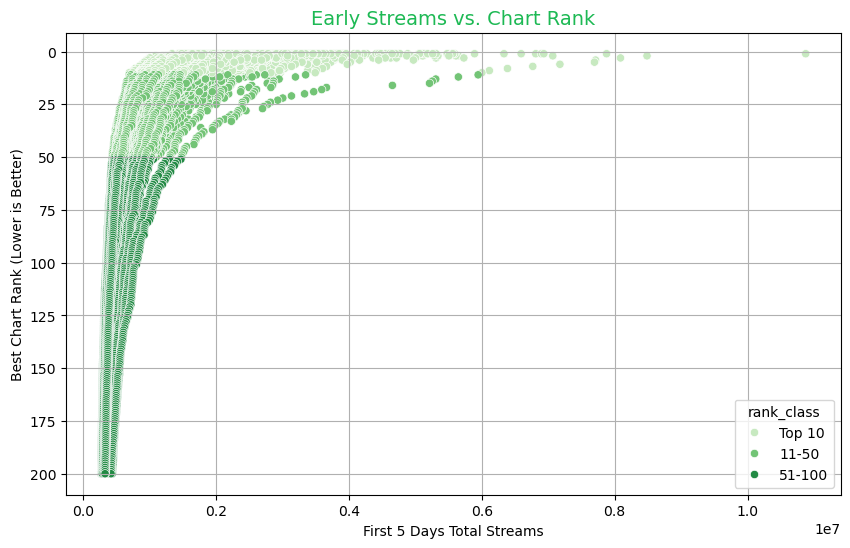}
\end{center}

This reinforced the hypothesis that while metadata adds power, it is not the only path to insight—some highly ranked songs didn’t have extreme stream counts early on. Thus, audio features could still provide predictive value in the absence of metadata.

\section{Modeling Methodology}

\subsection{Classifier Selection and Justification}

We evaluated four classification models:

\begin{enumerate}
    \item \textbf{Logistic Regression} served as a baseline due to its simplicity and interpretability.
    \item \textbf{K-Nearest Neighbors (KNN)} offered a non-parametric model that could capture local patterns in feature space.
    \item \textbf{Random Forest} was chosen for its robustness to outliers, ability to handle mixed feature types, and useful feature importance metrics.
    \item \textbf{XGBoost} was selected as a top-performing gradient boosting algorithm known for its efficiency and predictive power \cite{Chen2016}.
\end{enumerate}

Each model was first trained and evaluated using a holdout test set. Based on feedback, we then applied 5-fold stratified cross-validation and recorded the mean and variance of macro F1-scores to evaluate model stability. For Random Forest, Iwe performed hyperparameter tuning using GridSearchCV from the Scikit-learn library \cite{ScikitLearn2024}.

The evaluation strategy prioritized \textbf{macro-averaged F1-score}, which weights each class equally and is more appropriate in cases with class imbalance—such as the smaller number of Top 10 tracks in this dataset.

\subsection{Feature Importance and Filtered Models}

Tree-based models were especially useful in producing feature importance rankings. In the full-feature model, streams, previous\_rank, and days\_on\_chart emerged as the strongest predictors. However, when we trained a model with only audio features, new variables rose to the top—such as valence, acousticness, and tempo.

\begin{center}
\includegraphics[width=1\linewidth]{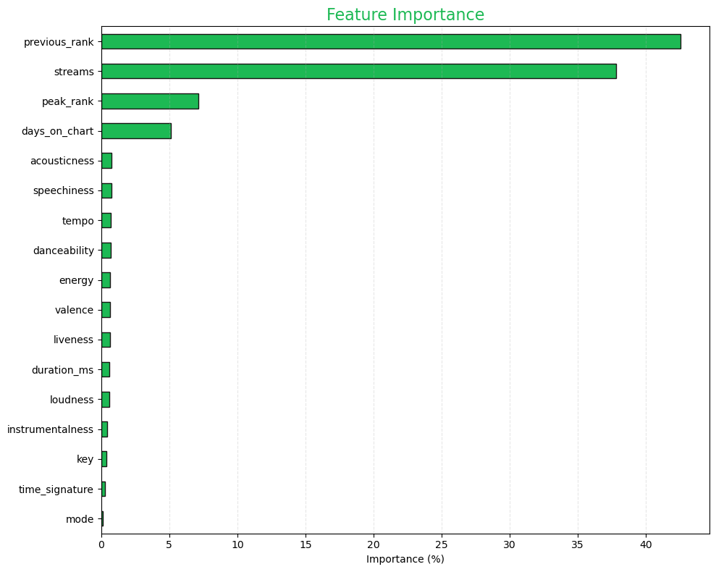}
\end{center}
\begin{center}
\includegraphics[width=1\linewidth]{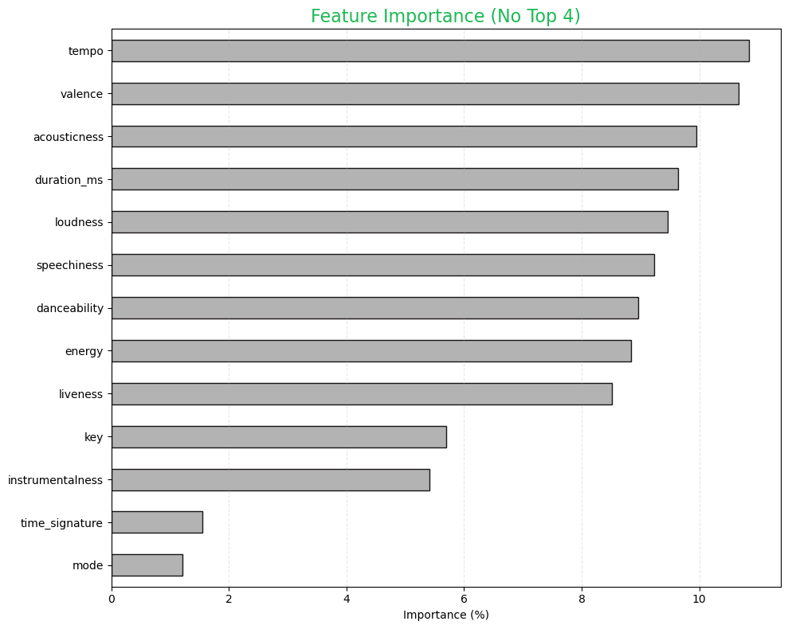}
\end{center}

This confirmed that audio alone can be informative. It also validated the decision to evaluate models both with and without metadata, since each version reveals different insights. The filtered model’s feature hierarchy suggests that upbeat, energetic songs (e.g., with high valence and tempo) may correlate with chart success even without prior exposure.

\section{RESULTS AND VISUALIZATIONS}

\subsection{Initial Holdout Performance (Before CV)}

The first round of modeling used a standard 80/20 train-test split. Table 1 summarizes the results.

\begin{tabular}{|p{3.4cm}|p{1.5cm}|p{1.7cm}|c|}
\hline
\textbf{Model} & \textbf{Accuracy} & \textbf{Macro F1} \\
\hline
Logistic Regression & 94\% & 0.90 \\
\hline
Random Forest & 97\% & 0.95 \\
\hline
K-Nearest Neighbors & 92\% & 0.85 \\
\hline
XGBoost & 97\% & 0.95 \\
\hline
\end{tabular}
		
Random Forest and XGBoost clearly outperformed other models. KNN struggled, likely due to its reliance on feature similarity—which was confounded by skewed distributions. Logistic Regression performed well overall but lacked the nuance to model interactions that tree-based models could capture.

The chart below visualizes the drop in performance when the four top metadata features (streams, previous\_rank, peak\_rank, days\_on\_chart) were excluded. Despite the reduction, Random Forest and XGBoost still achieved close to 89\% accuracy using only audio features.

\begin{center}
\includegraphics[width=1\linewidth]{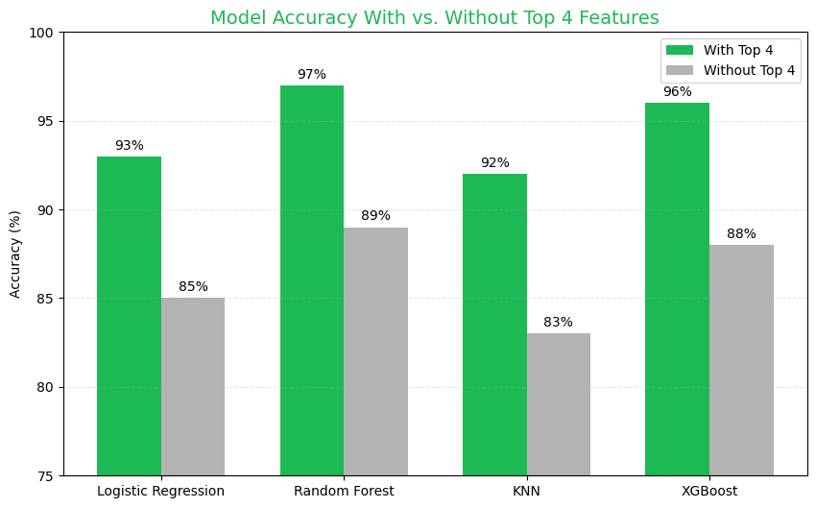}
\end{center}

This reinforced the value of musical structure in predicting popularity—useful for assessing unreleased tracks that don’t yet have metadata.

\subsection{Cross-Validation and Hyperparameter Tuning (Post-Feedback)}

Following reviewer feedback, we expanded model evaluation to include 5-fold stratified cross-validation. This approach helped capture performance variance across multiple data splits, providing a more reliable assessment of generalizability.

The table below shows the mean macro F1-score and standard deviation across folds:

\medskip
\begin{tabular}{|p{3cm}|p{4.3cm}|c|}
\hline
\textbf{Model} & \textbf{CV Macro F1 (Mean ± Std)} \\
\hline
Logistic Regression  & 0.8606 ± 0.0127 \\
\hline
Random Forest & 0.9434 ± 0.0028 \\
\hline
KNN  & 0.8489 ± 0.0050 \\
\hline
XGBoost & 0.9463 ± 0.0021 \\
\hline
\end{tabular}

Both Random Forest and XGBoost maintained high macro F1-scores and exhibited \textbf{very low standard deviation}, suggesting consistent performance regardless of data split. The stability of XGBoost in particular makes it highly deployable for real-world settings where prediction consistency is critical.

To further enhance model performance, we tuned Random Forest using GridSearchCV \cite{ScikitLearn2024}, searching across:
\begin{itemize}
    \item n\_estimators: 150
    \item max\_depth: None
    \item min\_samples\_split: 5
\end{itemize}

The best-performing configuration used n\_estimators = 150, max\_depth = None, and min\_samples\_split = 5. This configuration yielded a \textbf{macro F1-score of 0.9451}, almost identical to XGBoost’s performance but with slightly simpler interpretability via feature splits.

\subsection{Per-Class Metrics and Error Distribution}

One concern raised in feedback was whether the models performed equally well across all three classes—or if accuracy was driven by the dominant class (51–100). To evaluate this, we examined \textbf{class-specific precision, recall, and F1-scores} for the best model: \textbf{XGBoost}.

\medskip
\begin{tabular}{|p{1.5cm}|p{1.5cm}|p{1.7cm}|p{1.5cm}|c|}
\hline
\textbf{Class} & \textbf{Precision} & \textbf{Recall} & \textbf{F1-Score} \\
\hline
Top 10 & 0.932 & 0.962 & 0.947 \\
\hline
11–50 & 0.927 & 0.935 & 0.931  \\
\hline
51–100 & 0.987 & 0.983 & 0.985  \\
\hline
\end{tabular}
	
The results showed \textbf{remarkable balance}. Precision was slightly lower for the Top 10 class, meaning the model occasionally predicted a hit where there wasn’t one. However, recall was very high—indicating that most actual hits were successfully captured. This is an acceptable and often desirable tradeoff in real-world applications: it’s better to lean towards the side of catching potential hits, especially in music scouting or Artist \& Repertoire (A\&R) evaluation.

\section{Insights \& Conclusion}

The results of this project confirm that Spotify chart success can be meaningfully predicted using early features—especially those tied to musical composition. While metadata like stream count and rank history naturally enhance performance, \textbf{audio-only models remained competitive}, offering accuracy above 89\% and well-balanced F1-scores.

Tree-based models proved to be the most effective. Random Forest offered consistent, interpretable results with quick training time, while XGBoost delivered slightly higher precision and recall with excellent cross-validation stability. The inclusion of class-wise metrics and CV variance helped ensure fairness and trustworthiness, which are important considerations for any high-impact prediction system.

These findings have several practical implications:

\begin{itemize}
    \item \textbf{Pre-release screening:} Audio-only models can be used before a song drops to estimate likelihood of success.
    \item \textbf{Post-release monitoring:} Combining early streams with musical features can help identify sleeper hits.
    \item \textbf{Playlist curation:} Class probability scores can help optimize inclusion decisions for editorial or algorithmic playlists.
\end{itemize}

This work shows that combining musicality with basic modeling techniques can yield powerful insights, bridging art and data in a meaningful, measurable way.

\section{Lessons Learned}

This project evolved significantly from its initial form. At first, we focused on traditional model training and test-set evaluation. But after receiving reviewer feedback, we gained a deeper understanding of what constitutes \textbf{robust and fair evaluation}. We learned to implement:

\begin{itemize}
    \item Cross-validation for stability
    \item Macro F1 for fairness across classes
    \item Hyperparameter tuning for optimization
    \item Class-wise analysis for bias detection
\end{itemize}

These additions transformed the project from a modeling exercise into a statistically grounded investigation. we also saw firsthand the importance of data preprocessing consistency, label encoding, and feature scaling—small details that can quietly derail or elevate a project.

Perhaps most meaningfully, this project reinforced the idea that data science is iterative. The best outcomes come from \textbf{combining technical skill with humility,} knowing that every model benefits from scrutiny, improvement, and context. We now carry this mindset forward into all our future work.


\bibliographystyle{ACM-Reference-Format}
\bibliography{references}

\end{document}